\documentclass[fleqn,usenatbib]{mnras}

\usepackage[T1]{fontenc}

\DeclareRobustCommand{\VAN}[3]{#2}
\let\VANthebibliography\thebibliography
\def\thebibliography{\DeclareRobustCommand{\VAN}[3]{##3}\VANthebibliography}

\usepackage{graphicx}	% Including figure files
\usepackage{amsmath}	% Advanced maths commands
\usepackage{amssymb}	% Extra maths symbols

\title[Thermal state of the intergalactic medium near to the optical limit for the Ly-$\alpha$ forest]{Thermal state of the intergalactic medium near to the optical limit for the Ly-$\alpha$ forest}

\author[T. Ondro and R. G\'{a}lis]{
Tom\'{a}\v{s} Ondro$^{1, 2}$\thanks{E-mail: tomas.ondro@mendelu.cz} and
Rudolf G\'{a}lis$^{2}$
\\
$^{1}$Department of Agricultural, Food and Environmental Engineering, Faculty of AgriSciences, Mendel University in Brno,\\ Zem\v{e}d\v{e}lsk\'{a} 1, 613 00 Brno, Czech Republic\\
$^{2}$Institute of Physics, Faculty of Science, Pavol Jozef \v{S}af\'{a}rik University in Ko\v{s}ice, Park Angelinum 9,
040 01 Ko\v{s}ice, Slovakia\\
}

\date{Accepted XXX. Received YYY; in original form ZZZ}

\pubyear{2020}
\usepackage{newtxtext,newtxmath,amsmath,amssymb}
\begin{document}
\label{firstpage}
\pagerange{\pageref{firstpage}--\pageref{lastpage}}
\maketitle

\begin{abstract}
In this article, the temperature-density relation of the intergalactic medium was studied in the region $1.6 \leq z < 2.0$ divided into two bins. For this purpose, the Ly-$\alpha$ forest decomposition into individual absorption profiles was used for the study of 35 publicly available quasar spectra obtained by the Ultraviolet and Visual Echelle Spectrograph (UVES) on the Very Large Telescope (ESO) and by the High Resolution Echelle Spectrometer (HIRES) on the Keck Telescope. For the determination of the thermal state sensitive cut-off position in the $b - N_{\ion{H}{i}}$ distribution, the iterative fitting procedure was adopted. The measurements were calibrated using mock Ly-$\alpha$ forest data generated by 23 hydrodynamical simulations with different thermal histories. The value of the temperature at mean density corresponds to the decreasing trend predicted by various models at the lower redshifts. In the case of power law index,  determined values are close to 1.6, which is expected after all reionization events in various models assuming the balance of photoheating with adiabatic cooling.
\end{abstract}

\begin{keywords}
cosmology: early Universe -- quasars: absorption lines -- intergalactic medium
\end{keywords}

%%%%%%%%%%%%%%%%%%%%%%%%%%%%%%%%%%%%%%%%%%%%%%%%%%

%%%%%%%%%%%%%%%%% BODY OF PAPER %%%%%%%%%%%%%%%%%%

\section{Introduction}
The physical conditions of the intergalactic medium (IGM) could be studied by analysis of the so-called Ly-$\alpha$ forest in quasar (QSO) spectra. \cite{HuiGnedin} showed that for gas overdensities $\delta \lesssim 10$, the temperature is related to the density throughout a power-law in the form
 \begin{equation}
T=T_{0}(1+\delta)^{\gamma-1},
\label{asymptot_td_relation2}
\end{equation}
where $T_{0}$ is the temperature of the IGM at the mean density and $\gamma$ is the power-law index. If we determine the parameters $T_{0}$ and $\gamma$ as a~function of redshift, we can describe the thermal history of the IGM.

Various approaches were used for the characterization of the $T-\rho$ relation of the IGM from Lyman-$\alpha$ absorption forest. As an example, the approach that treats the Ly-$\alpha$ forest as a superposition of discrete absorption profiles \citep{Schaye1999,Ricotti2000,Rorai2018,Hiss_2018}, analysis of the flux probability distribution function by \cite{Bolton2008} and \cite{Viel2009}, the power spectrum of the transmitted flux \citep{Zaldarriaga2001,Theuns2002,Walther2019}, the average local curvature \citep{Becker2011} and \cite{Boera2014} and wavelet decomposition \citep{Lidz2010}.

Most studies deal with the characterization of the $T-\rho$ relation of the IGM around $z \sim 3$, which is interesting due to the He\,{\sc II} reionization phase. Unlike of this redshift, for which is possible to find $\sim300$ Ly-$\alpha$ absorption lines in QSO spectra, for $z\sim2$, there are typically less than 100 absorption lines with $\log{N_{\text{H\,{\sc i}}}} \in (12.5 - 14.5)$ between the Ly-$\alpha$ and Ly-$\beta$ emission lines \citep{Schaye1999}. Generally, the number density of the Ly-$\alpha$ forest decreases with the decreasing redshift, and $z \simeq 1.5$ is considered as the optical limit for the Ly-$\alpha$ forest \citep{Boera2014}.

The temperature in this lower redshift region was firstly studied by \citet{Boera2014} using the curvature measurement. For the mean redshift $z = 1.63$, the authors determined the temperature $(20.66 \pm 2.03)\times 10^{3}$\,K and $(13.00 \pm 1.27) \times 10^{3}$\,K at the mean density assuming the values of $\gamma \sim$ 1.3 and 1.5, respectively. For the higher value of mean redshift $(z = 1.82)$, the authors determined the values of teplerature $T_{0}(\gamma\sim1.3) = (17.61 \pm 0.76) \times 10^{3}$\,K and $T_{0}(\gamma\sim1.5) = (11.79 \pm 0.51) \times 10^{3}$\,K. Based on the results from the redshift range $1.5 \lesssim z \lesssim 2.8$, the authors of this study found a decrease in temperature with decreasing redshift, which could be interpreted as a trace of completion of the reheating process related to the \ion{He}{ii} reionization.

\citet{Walther2019} presented the fiducial evolution of thermal parameters in the region $1.8 < z <5.4$ using the Ly-$\alpha$ flux power spectrum. Assuming the strong prior on mean transmitted flux $\overline{F}$, the corresponding values of the temperature $T_{0}$ and power-law index $\gamma$ are $(7.68^{+3.69}_{-2.18}) \times 10^{3}$\,K and $1.63^{+0.16}_{-0.25}$ at the redshift $z = 1.8$, respectively.

It is worth noting that \citet{Schaye2000} also studied the region $1.85 \leq z \leq 2.09$ using the spectrum of only one quasar (Q1100-264). For the median value of the redshift $z = 1.96$ , the authors determined value of $T_{0} \sim 11\,000$\,K and $\gamma = 1.4$. In this case, the analysis of IGM was based on the Voigt profile decomposition of the Ly-$\alpha$ forest into the set of individual absorption lines. 

The aim of this study is to determine the coefficients in the $T-\rho$ relation of the IGM in the redshift range of $1.6 \leq z < 2.0$. The article is organized as follows: In Section \ref{sec:data_analysis} we present the basic information about the used data and the analysis of the spectra based on the Voigt profile fitting. This section also contains the description of the metal line and narrow line rejection, as well as the cut-off fitting procedure. The information about the hydrodynamical simulations is given in Section \ref{sec:sim}. The {\sc THERMAL} suite (Thermal History and Evolution in Reionization Models of Absorption Lines) is also introduce in this section. A description of the determination of coefficients in the $T-\rho$ relation and the calibration procedure are presented in Section \ref{sec:T_rho_rel}. In Section \ref{sec:results} we present our results and their comparison with the previously published ones and with simulations. Finally, our conclusions are given in Section \ref{sec:conclusions}.

\begin{table}
	\centering
	\caption{List of QSOs whose spectra were used in this study. The S/N ratio was calculated according to \citet{Stoehr2008} for the spectral regions where the absorbers were parameterized.}
	\label{tab:QSO_data}
	\scriptsize
	\begin{tabular}{lllccl} 
		\hline
		\textbf{Object} & \textbf{$\alpha_{2000}$} & \textbf{$\delta_{2000}$} & \textbf{\boldmath{$z_{\rm{em}}$}} & \textbf{S/N} & Source\\
		 & [h m s] & [d m s] &  &  & \\
		\hline
J000344-232355	&	00 03 44.91	&	--23 23 55.3	&	2.28	&	49	&	UVES	\\
J000443-555044	&	00 04 43.28	&	--55 50 44.6	&	2.1	    &	10	&	UVES	\\
J001602-001225	&	00 16 02.40	&	--00 12 25.1	&	2.085	&	18	&	UVES	\\
J012417-374423	&	01 24 17.36	&	--37 44 23.0	&	2.2	    &	37	&	UVES	\\
J014333-391700	&	01 43 33.63	&	--39 17 00.1	&	1.807	&	30	&	UVES	\\
J014944+150106	&	01 49 44.43	&	15 01 06.70	    &	2.06	&	12	&	KODIAQ	\\
J022620-285750	&	02 26 20.49	&	--28 57 50.7	&	2.171	&	10	&	UVES	\\
J022853-033737	&	02 28 53.21	&	--03 37 37.1	&	2.066	&	11	&	KODIAQ	\\
J024008-230915	&	02 40 08.16	&	--23 09 15.6	&	2.223	&	83	&	UVES	\\
J025634-401300	&	02 56 34.00	&	--40 13 00.3	&	2.29	&	10	&	UVES	\\
J042707-130253	&	04 27 07.29	&	--13 02 53.6	&	2.159	&	10	&	UVES	\\
J092129-261843	&	09 21 29.34	&	--26 18 43.2	&	2.3	    &	14	&	UVES	\\
J101939+524627	&	10 19 39.15	&	52 46 27.80	    &	2.17	&	10	&	KODIAQ	\\
J103921-271916	&	10 39 21.84	&	--27 19 16.4	&	2.23	&	25	&	UVES	\\
J110325-264515	&	11 03 25.29	&	--26 45 15.8	&	2.145	&	61	&	UVES	\\
J110610+640009	&	11 06 10.74	&	64 00 09.60	    &	2.203	&	29	&	KODIAQ	\\
J121140+103002	&	12 11 40.59	&	10 30 02.0	    &	2.191	&	11	&	UVES	\\
J122824+312837	&	12 28 24.96	&	31 28 37.60	    &	2.2	    &	46	&	KODIAQ	\\
J124913-055919	&	12 49 13.86	&	--05 59 19.1	&	2.247	&	13	&	UVES	\\
J124924-023339	&	12 49 24.86	&	--02 33 39.7	&	2.12	&	11	&	UVES	\\
J131011+460124	&	13 10 11.61	&	46 01 24.50	    &	2.133	&	12	&	KODIAQ	\\
J133335+164903	&	13 33 35.78	&	16 49 03.9	    &	2.089	&	52	&	UVES	\\
J134427-103541	&	13 44 27.06	&	--10 35 41.7	&	2.134	&	47	&	UVES	\\
J141719+413237	&	14 17 19.23	&	41 32 37.00	    &	2.024	&	12	&	KODIAQ	\\
J141906+592312	&	14 19 06.31	&	59 23 12.20	    &	2.321	&	18	&	KODIAQ	\\
J144653+011355	&	14 46 53.05	&	01 13 55.9	    &	2.216	&	17	&	UVES	\\
J145102-232930	&	14 51 02.49	&	--23 29 30.9	&	2.215	&	53	&	UVES	\\
J152156+520238	&	15 21 56.48	&	52 02 38.40	    &	2.208	&	36	&	KODIAQ	\\
J162645+642655	&	16 26 45.69	&	64 26 55.20	    &	2.32	&	20	&	KODIAQ	\\
J212329-005052	&	21 23 29.46	&	--00 50 52.9	&	2.262	&	28	&	KODIAQ	\\
J221531-174408	&	22 15 31.65	&	--17 44 08.2	&	2.217	&	19	&	UVES	\\
J222756-224302	&	22 27 56.92	&	--22 43 02.5	&	1.891	&	32	&	UVES	\\
J225719-100104	&	22 57 19.04	&	--10 01 04.7	&	2.08	&	12	&	UVES	\\
J231324+003444	&	23 13 24.45	&	00 34 44.50	    &	2.083	&	11	&	KODIAQ	\\
J234023-005327	&	23 40 23.66	&	--00 53 27.0	&	2.085	&	18	&	KODIAQ	\\
		\hline
	\end{tabular}
\end{table}
%Ak myslite, že písmo nie je príliš malé, nechajme tu. Prípadne sa vyjadrí recenzent.

\section{Data and analysis}
\label{sec:data_analysis}
In this study, the sample of 35 publicly available quasar spectra were used for analysis. The first part of the sample consists of 13 QSO spectra from KODIAQ (Keck Observatory Database of Ionized Absorption toward Quasars) survey \citep{KodiaqDR,KodiaqDR1,KodiaqDR2}, which is a repository of reduced, continuum normalized spectra obtained by the High-Resolution Echelle Spectrometer (HIRES) on the Keck Telescope. These spectra were observed between years 1995--2012 and subsequently uniformly reduced and continuum fitted by eye by KODIAQ team \citep{KodiaqDR1}. The second part of the sample contains of 22 QSO spectra obtained by the Ultraviolet and Visual Echelle Spectrograph (UVES) on the VLT/ESO \citep{UVES_dataset}. The UVES Spectral Quasar Absorption Database comprises fully reduced, continuum-fitted high-resolution spectra of quasars in the redshift range $0 < z < 5$. From the whole dataset we selected only spectra that meet the following criteria:
\begin{enumerate}
\item the QSO spectrum covers at least $\Delta z $ = 0.1 of at least one considered redshift bin, 
\item the spectrum does not contain large gaps that would not allow the correct identification of metal lines,
\item the signal-to-noise (S/N) ratio of the spectrum is higher than 10 in the studied spectral region.
\end{enumerate} 

\begin{figure}
\centering
\includegraphics[width=\columnwidth]{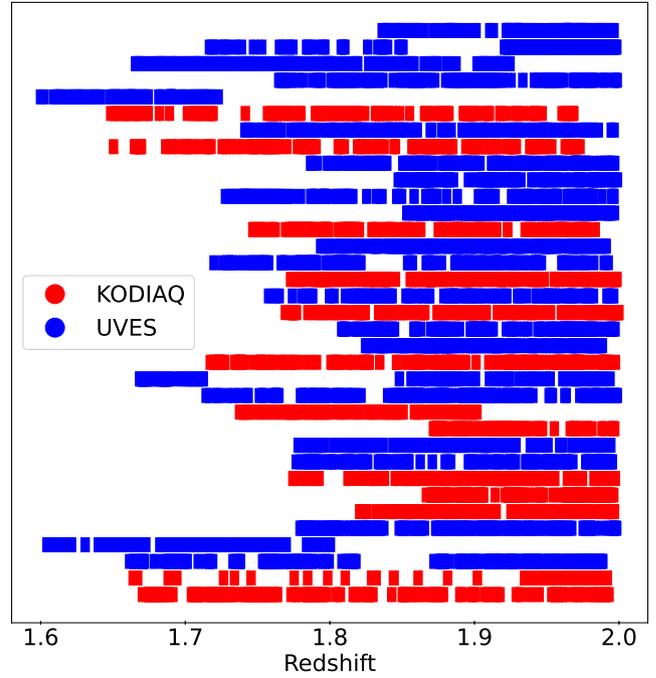}
\caption{The redshift coverage of the dataset used in this study.}
\label{fig:redshift_coverage}
\end{figure}

The list of 35 QSOs whose spectra were used in this study with their basic characteristics is presented in Tab. \ref{tab:QSO_data}. The redshift coverage of the analysed QSO spectra is shown in Fig. \ref{fig:redshift_coverage}. We studied the effects of continuum misplacement on the achieved results and showed its effect is negligible in the case of our data. A detailed description of the analysis can be found in Appendix \ref{continuum_misplacement}. 
\subsection{Voigt profile fitting}
\label{sec:VPFIT}
In this study, the {\sc VPFIT} code written by R.~F.~Carswell and J.~K.~Webb \citep{Carswell2014} was used. The QSO spectra were analyzed within the following parameter space: $b = 1 - 300\,\rm km\,s^{-1}$ and $\log{N_{\ion{H}{i}}} = 11.5 - 16.0$. The rest-frame wavelengths 1050 -- 1180\,\AA\, inside the Ly-$\alpha$ forest were chosen for the Voigt profile fitting to avoid proximity effects, i.e. a region affected by the local QSO radiation (exclude the Ly-$\beta$ and O\,{\sc VI} $\lambda 1035$ emission lines) rather than the metagalactic ultraviolet background \citep[UVB;][]{Palanque2013,Walther2018,Hiss_2018}.

\begin{figure*}
\centering
\includegraphics[width=0.9\linewidth]{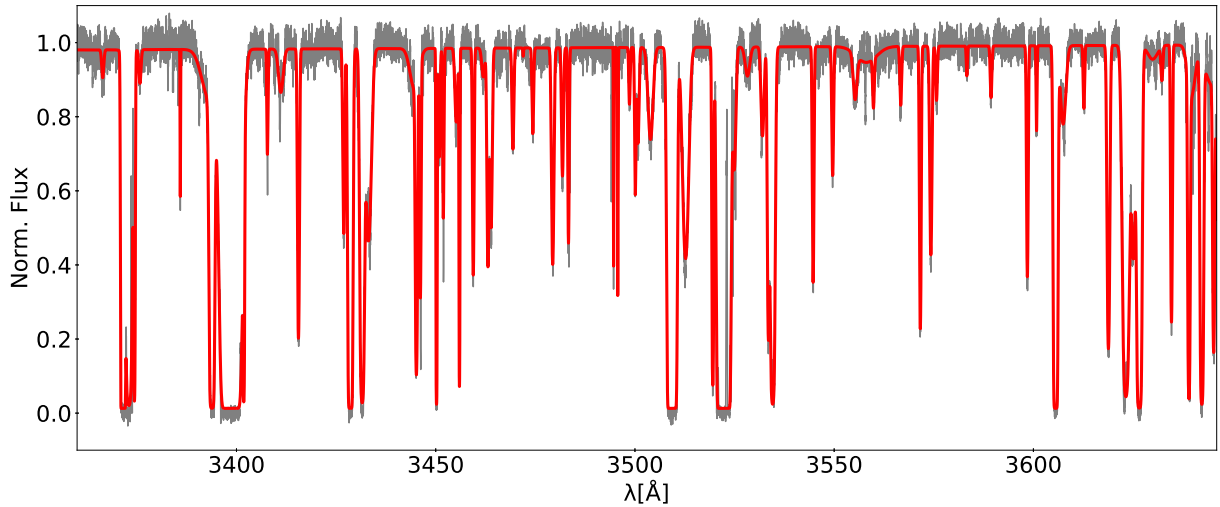}
\caption{Resulting fit of the Ly-$\alpha$ forest in the spectrum of QSO J122824+312837. The original spectrum (grey line) is well described by the superposition of Voigt profiles fitted using the {\sc VPFIT} code (red line).}
\label{Abs_lines}
\end{figure*}

The Ly-$\alpha$ absorbers corresponding to the damped Ly-$\alpha$ (DLA) systems with $\log{N_{\ion{H}{i}}} \gtrsim 20$ and also sub-damped Ly-$\alpha$ (sub-DLA) systems were identified by eye and excluded from the analysis. Each spectrum was also visually observed for bad points and gaps, which were subsequently masked and cubically interpolated. Relative errors of 1\% were given to such spectral regions, so the Voigt profile fitting procedure was not influenced \citep{Hiss_2018}. An example of the decomposition of the Ly-$\alpha$ forest into individual absorption lines in the spectrum of QSO J122824+312837 is shown in Fig. \ref{Abs_lines}.
\subsection{Metal lines rejection}
Quasar spectra also contain various metal absorption lines, which can affect the obtained results. They are usually related to the strong \ion{H}{i} absorption, and can be identified via associations with other ionic metal lines. From this reason, we identified DLA and sub-DLA systems and determined their redshifts with help of the associated \ion{C}{iv}, \ion{Si}{iv}, \ion{Mg}{ii} metal lines redward of the Ly-$\alpha$ emission peak. It is worth noting that in the case of metal absorption lines not related with DLA systems (mostly within the data coverage), we used the doublets of \ion{Si}{iv}, \ion{C}{iv} and \ion{Mg}{ii} to determine the redshift of metal absorption systems. If the redshifts were known, the other metal lines (see Tab. \ref{tab:Metal_transitions}) were determined based on their characteristic $\Delta\lambda$.

\begin{figure}
\centering
	\includegraphics[width=\columnwidth]{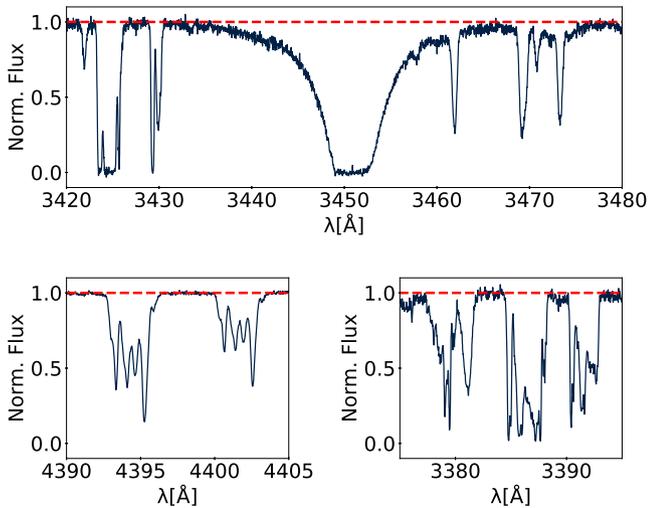}
 \caption{
Example of the metal lines rejection procedure based on the sub-DLA system at $z \approx 1.839$ in the spectrum of QSO J110325-264515 (upper panel). The associated \ion{C}{iv} 1548, 1550\,\AA\, spectral lines redward of the Ly-$\alpha$ emission peak (bottom left panel) were used for more precise determination of the sub-DLA redshift. The \ion{Si}{ii} 1190, 1193\,\AA\, spectral lines (bottom right panel) are related to the strong \ion{H}{i} absorption depicted in upper panel. The red dashed lines represent the continuum level.}
    \label{fig:metal_rejection}
\end{figure}

After the identification, the procedure for the metal line rejection in both cases can be, in general, described as follows: firstly, we fitted the studied intervals unmasked of identified lines to minimise the impact of possible adverse effects of masking on the Voigt profile fitting. After this step, we excluded the previously identified metal lines from the {\sc VPFIT} output so that they do not affect the results of our analysis. Example of the metal lines rejection procedure based on the sub-DLA system at $z \approx 1.839$ in the spectrum of QSO J110325-264515 is shown in Fig. \ref{fig:metal_rejection}.

As was described by \citet{Hiss_2018}, we also tested if the remaining absorbers below the lower envelope of the $b-N_{\text{H\,{\sc I}}}$ distribution belong to the metal lines listed in Tab. \ref{tab:Metal_transitions}. We would like to note that if the absorber candidate was identified only as a single or doubtful doublet feature, we did not consider it as a metal line. 

\begin{table}
	\begin{center}
	\caption{List of identified metal lines with their oscillator strength $f$.}
	\label{tab:Metal_transitions}
	\begin{tabular}{lccc} 
		\hline
		\textbf{Absorber} & \textbf{$\lambda_{\text{rest}}$ [\AA]} & \textbf{$f$} & \textbf{Reference} \\
		\hline
\ion{O}{VI}	&	1031.9261	&	0.13290	&	1	\\
\ion{C}{II}	&	1036.3367	&	0.12310	&	1	\\
\ion{O}{VI}	&	1037.6167	&	0.06609	&	1	\\
\ion{N}{ii}	&	1083.9900	&	0.10310	&	1	\\
\ion{Fe}{iii}	&	1122.5260	&	0.16200	&	2	\\
\ion{Fe}{ii}	&	1144.9379	&	0.10600	&	3	\\
\ion{Si}{ii}	&	1190.4158	&	0.25020	&	1	\\
\ion{Si}{ii}	&	1193.2897	&	0.49910	&	1	\\
\ion{N}{i}	&	1200.2233	&	0.08849	&	1	\\
\ion{Si}{iii}	&	1206.5000	&	1.66000	&	1	\\
\ion{N}{v}	&	1238.8210	&	0.15700	&	1	\\
\ion{N}{v}	&	1242.8040	&	0.07823	&	1	\\
\ion{Si}{ii}	&	1260.4221	&	1.00700	&	1	\\
\ion{O}{i}	&	1302.1685	&	0.04887	&	1	\\
\ion{Si}{ii}	&	1304.3702	&	0.09400	&	4	\\
\ion{C}{ii}	&	1334.5323	&	0.12780	&	1	\\
\ion{C}{ii}*	&	1335.7077	&	0.11490	&	1	\\
\ion{Si}{iv}	&	1393.7550	&	0.52800	&	1	\\
\ion{Si}{iv}	&	1402.7700	&	0.26200	&	1	\\
\ion{Si}{ii}	&	1526.7066	&	0.12700	&	5	\\
\ion{C}{iv}	&	1548.1950	&	0.19080	&	1	\\
\ion{C}{iv}	&	1550.7700	&	0.09522	&	1	\\
\ion{Al}{ii}	&	1670.7874	&	1.88000	&	1	\\
\ion{Al}{iii}	&	1854.7164	&	0.53900	&	1	\\
\ion{Al}{iii}	&	1862.7895	&	0.26800	&	1	\\
\ion{Mg}{ii}	&	2796.3520	&	0.61230	&	6	\\
\ion{Mg}{ii}	&	2803.5310	&	0.30540	&	6	\\
		\hline
	\end{tabular}\\
	\end{center}
References: (1) \cite{Morton1991}, (2) \cite{Prochaska2001}, (3) \cite{Howk2000}, (4) \cite{Tripp1996}, (5) \cite{Schectman1998}, (6) \cite{Verner1996}.
\end{table}

\subsection{The \boldmath{$b-N_{\ion{H}{i}}$} distributions}
\begin{figure}
\centering
	\includegraphics[width=\columnwidth]{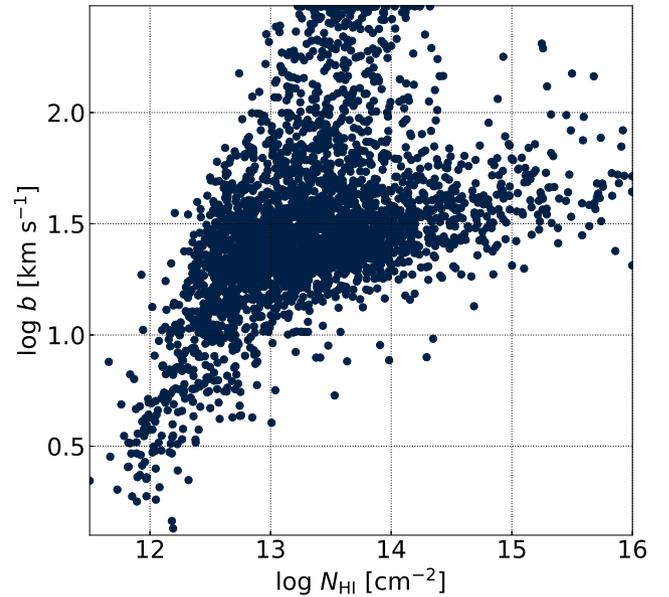}
    \caption{The $b-N_{\text{H\,{\sc I}}}$ distribution of the parameterized absorbers in the redshift range of $1.6 \leq z < 2.0$ with relative errors in the Doppler widths and column densities less than 50\%.}
    \label{fig:bN_distr}
\end{figure}

The {\sc VPFIT} outputs were used to generate the $b-N_{\text{H\,{\sc I}}}$ distributions, from which parameters in the $T-\rho$ relation could be determined. The $b-N_{\text{H\,{\sc I}}}$ distribution of the parameterized absorbers in the redshift range of $1.6 \leq z < 2.0$ is plotted in Fig. \ref{fig:bN_distr}. Firstly, we divided the results into two bins with the redshift range of $\langle 1.6 - 1.8)$, and $\langle 1.8 - 2.0)$. Then, the lower limit of $b$ was chosen to $10$\,km\,s$^{-1}$ mainly due to the lines with lower values could be attributed to the ionic metal-line transitions \citep{Hiss_2018,Rauch_1997}. The lines with $b > 100$ \,km\,s$^{-1}$ were also excluded, which is a convention used for example by \citet{Rudie2012} and \citet{Hiss_2018}. The reason for such a choice is that the turbulent broadening dominates over the thermal one in these lines. In the case of the column density, the chosen range of $12.5 \leq \log{N_{\ion{H}{i}}} \leq 14.5$ corresponds to the gas density range for which the equation of state is well fitted by a power law \citep{Schaye1999}.

\subsection{Narrow lines rejection}
Although the aforementioned rejection procedure was applied, many narrow absorption lines in blends and unidentified metal lines with $b \geq 10 \,\rm km\,s^{-1}$ still remain in the results. Due to this fact, we used the similar iterative rejection algorithm as proposed by \citet{Hiss_2018}, which can be described as follows: firstly, we divided the absorbers with $b < 40\,\rm km s^{-1}$ into eight $\log{N_{\ion{H}{i}}}$ bins of the same size. Then, the mean and standard deviation of the absorbers in every bin were calculated. After this step, we excluded all points below $2\sigma$ of the mean. This procedure is iterated until no points are excluded. After the last iteration, we fitted the line to the $\log{b_{2\sigma}}$ values of each $\log{N_{\ion{H}{i}}}$ bin. Finally, if the position of this line is determined, all absorbers below it are excluded. 
\begin{figure*}
\centering
	\includegraphics[width=0.7\columnwidth]{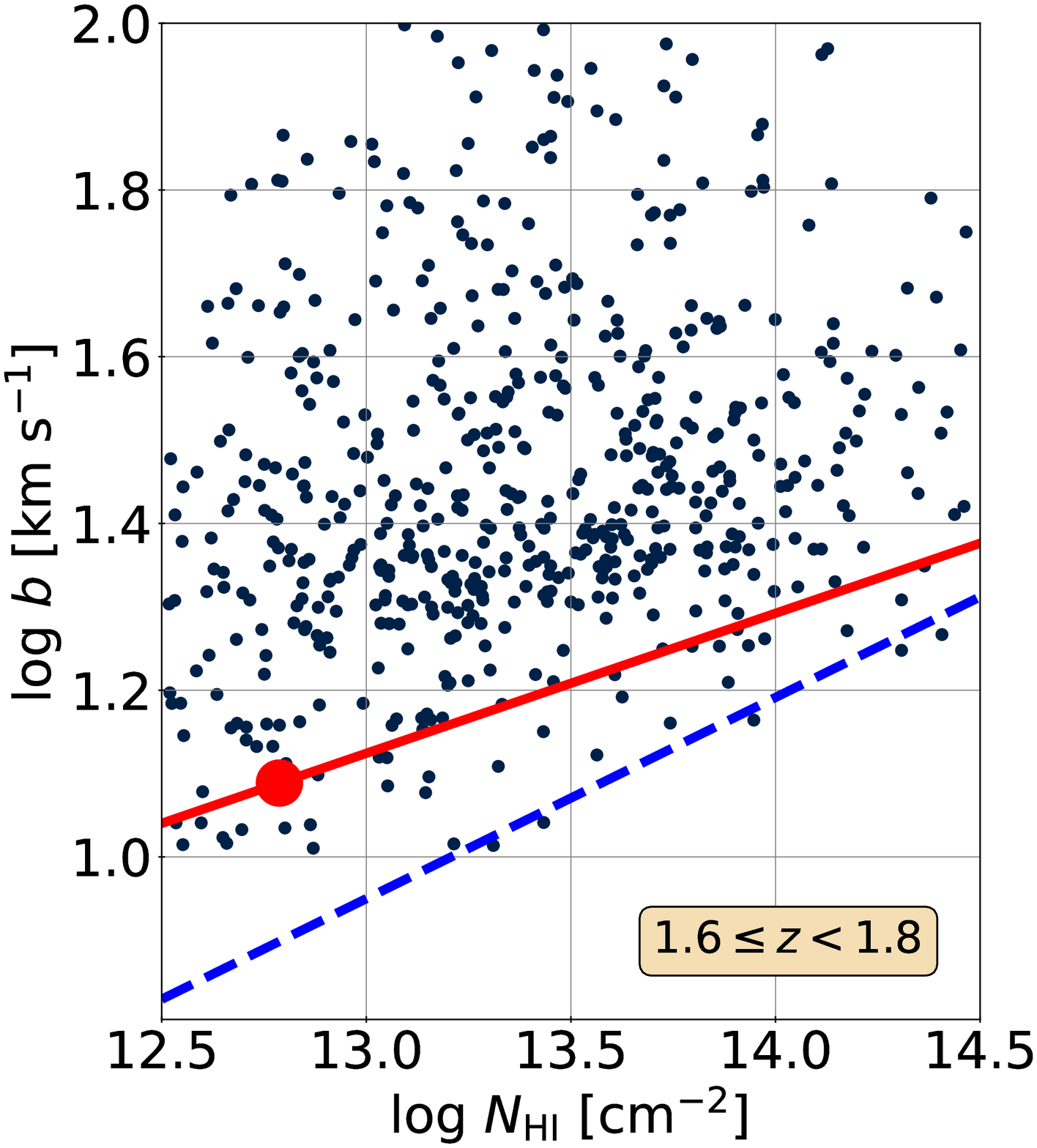} \hspace{3em}
	\includegraphics[width=0.655\columnwidth]{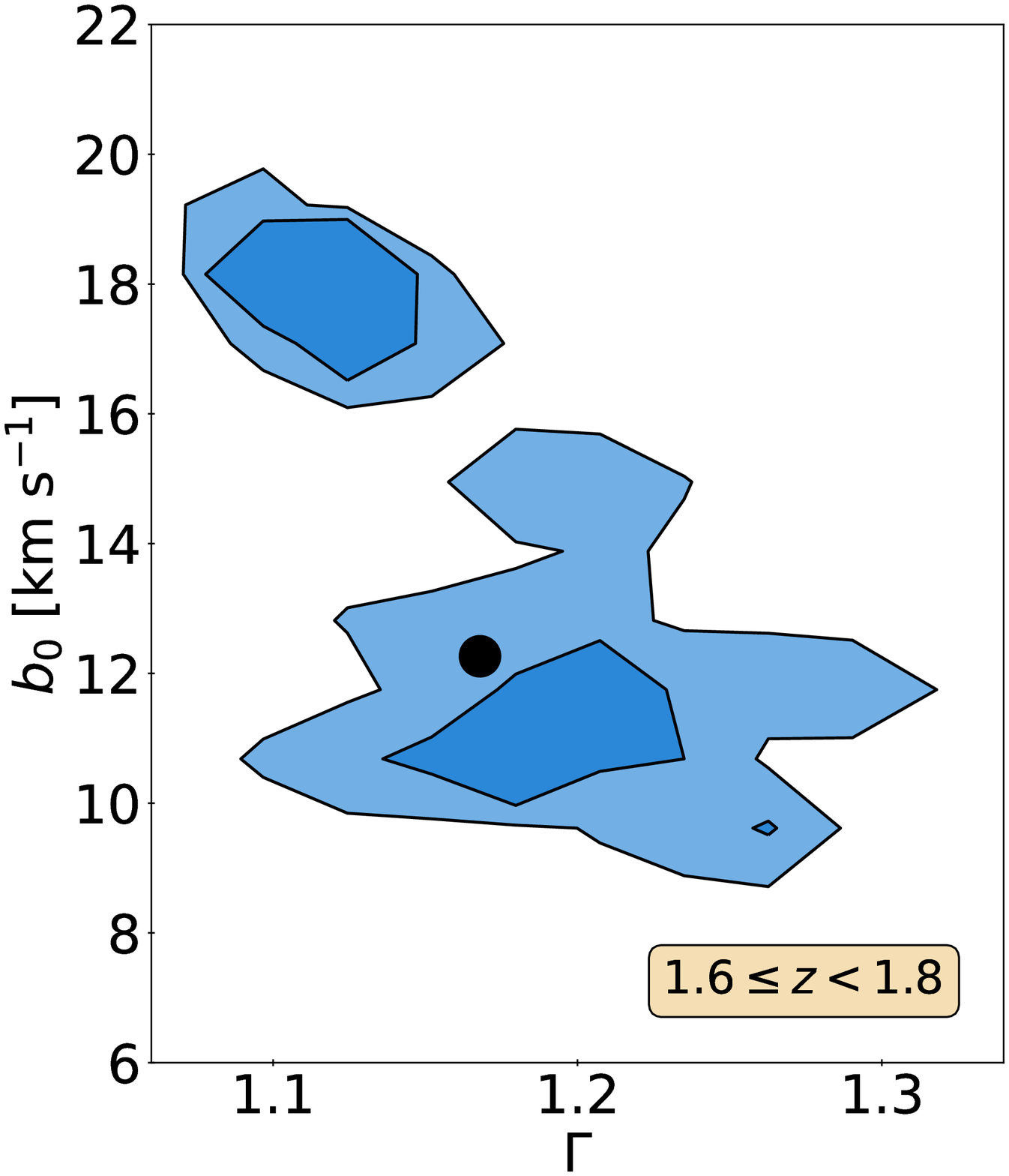} \\
	\vspace{2em}
	\includegraphics[width=0.7\columnwidth]{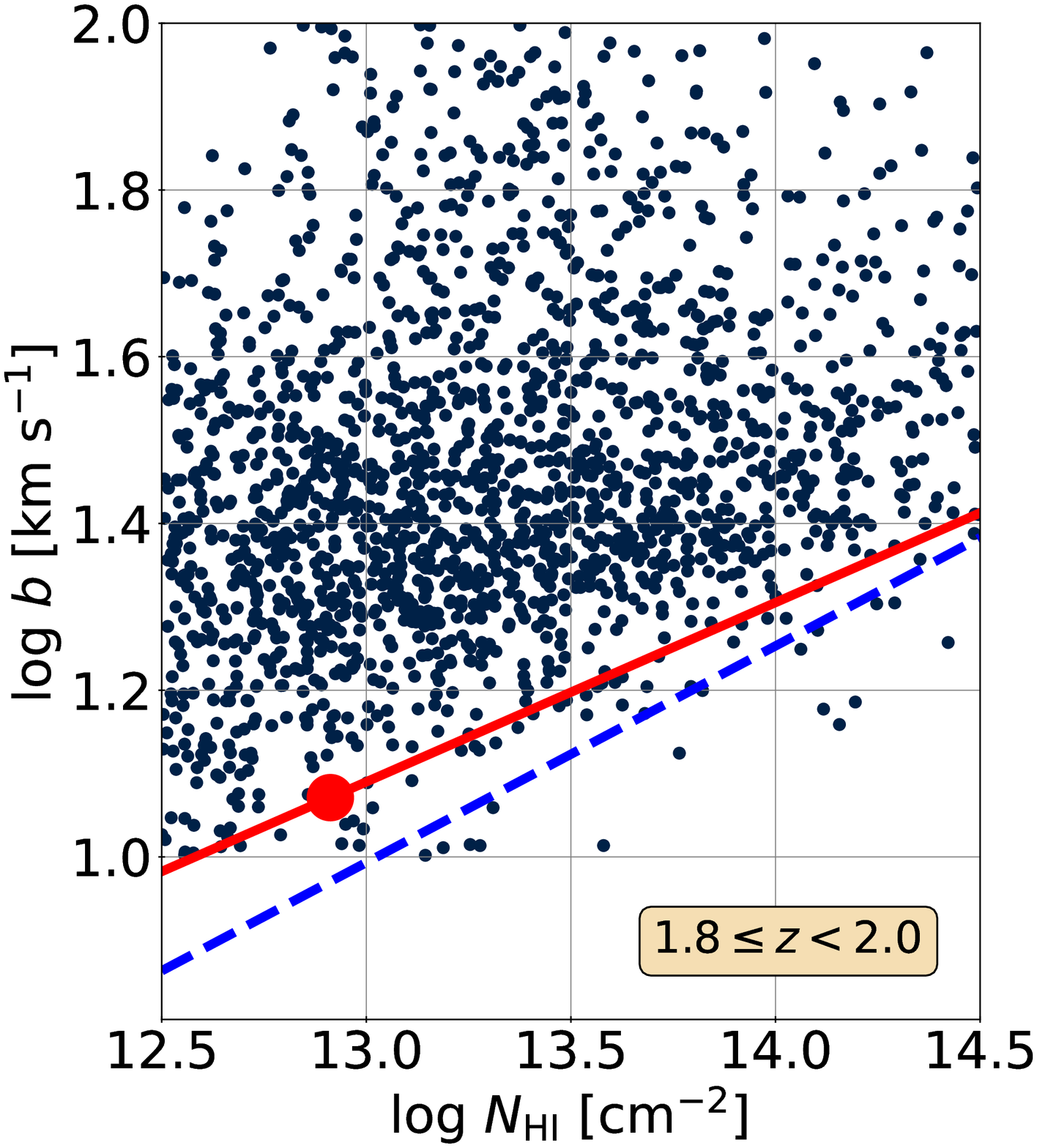}\hspace{3em}
	\includegraphics[width=0.655\columnwidth]{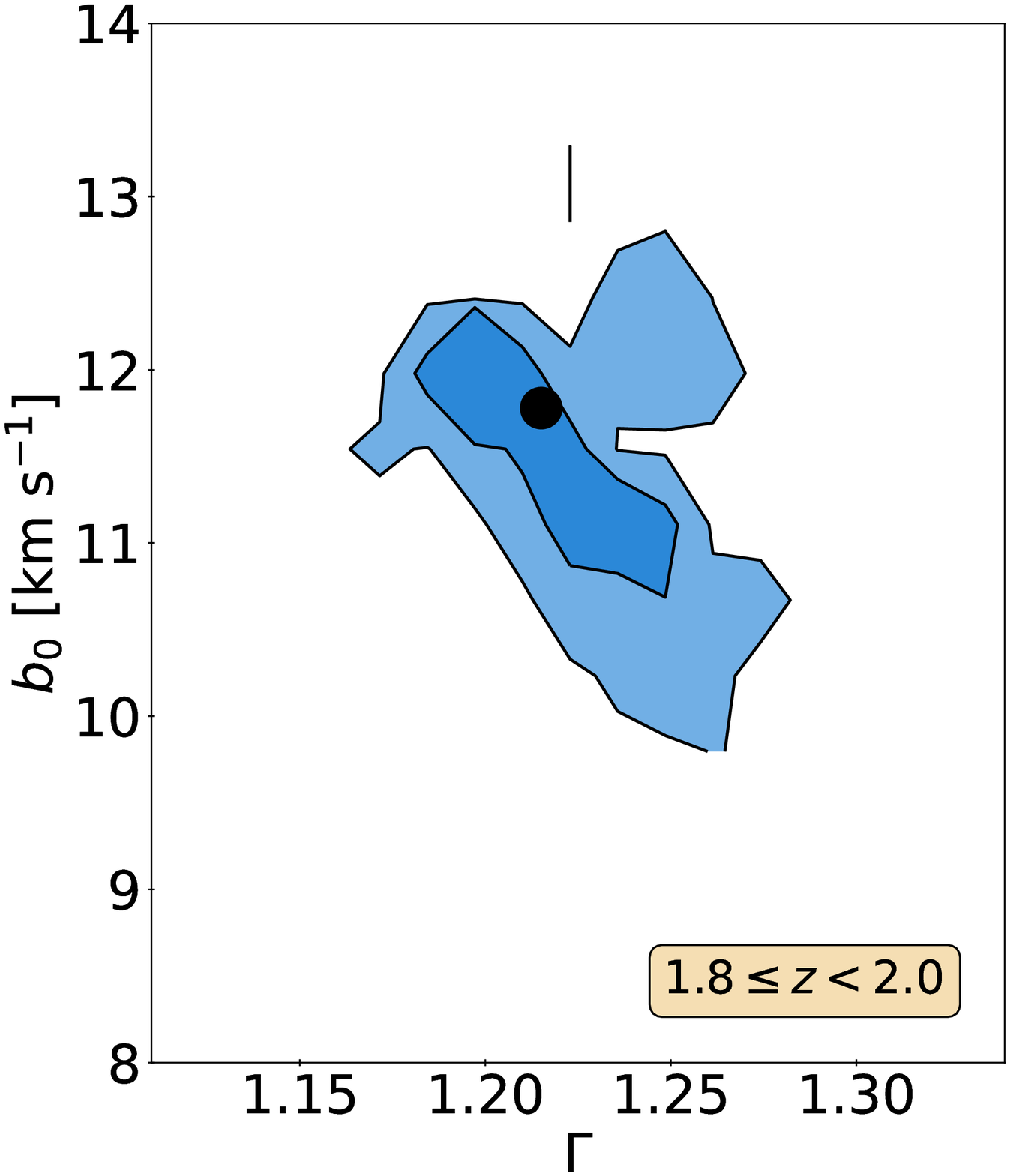}
    \caption{$b - N_{\ion{H}{i}}$ distributions are shown in left panels, where the red solid and blue dashed lines correspond to the best cut-off fits and the results of narrow-line rejection procedure, respectively. The overplotted red circles correspond to the $\log{N_{\ion{H}{i},0}}$. The right panels show the corresponding PDFs $p(b_{0}, \Gamma)$ , where the 68\% and 95\% confidence levels are plotted by dark and light blue colors, respectively. The black points correspond to the medians of the marginal distributions of $b_{0}$ and $\Gamma$.}
    \label{fig:cutoff_results}
\end{figure*}

\subsection{Fitting the cut-off in the \boldmath{$b-N_{\ion{H}{i}}$} distribution}
\label{Sec:cut_off}
In this study, the iterative fitting procedure proposed by \citet{Schaye1999} was used to determine the position of the thermal state sensitive cut-off. It is based on the following equation
\begin{equation}
\log{b_{\rm th}}=\log{b_{0}}+(\Gamma-1)\log{N_{\ion{H}{i}}/N_{\ion{H}{i},0}}.
\label{cutoff1}
\end{equation}
where $b_{\rm th}$ corresponds to the thermal Doppler broadening, $b_{0}$ is the minimal broadening value at column density $N_{\ion{H}{i},0}$ and $\Gamma$ is the index of this power-law relation. Firstly, we fit the Eq. (\ref{cutoff1}) to the points in the $b-N_{\ion{H}{i}}$ distribution using the least absolute deviations method. After this step, the mean absolute deviation is calculated as
\begin{equation}
|\delta \log{b}| = \frac{1}{n}\sum_{i=1}^{n}|\log{b_{i}} - \log{b_{\rm th}(N_{\ion{H}{i},\rm i})|}
\label{cutoff2}
\end{equation}
where $n$ is the sample size. Subsequently, all points for which the value of the Doppler parameter $b$ fulfills the condition
\begin{equation}
\log{b}>\log{b_{\rm th}} + |\delta \log{b}|
\label{cutoff3}
\end{equation}
are excluded in the next iteration. All these steps are repeated until no points are more than one absolute mean deviation above the fit. The last step is the exclusion of the points that are below the one absolute mean deviation of the last fit. The remaining points are then used for the final fit, from which $b_{0}$ and $(\Gamma-1)$ are determined.

\subsection{Estimation of the $N_{\ion{H}{i},0}$}
As we mentioned before, the value of $N_{\ion{H}{i},0}$ is just a normalization. However, it is useful to choose its value that corresponds to the column density of a typical absorber at the mean density of the IGM \citep{Hiss_2018}. Under the assumption of the local hydrostatic equilibrium in the low-density cloud, the relationship between $N_{\ion{H}{I}}$ and local overdensity $\Delta = \rho/\overline{\rho}$ can be written as \citep{Shaye2001,Rudie2012} 

\begin{equation}
N_{\ion{H}{i},0} \simeq 10^{13.23} \Delta^{3/2} \frac{T_{4}^{-0.22}}{\Gamma_{\text{ion},\ion{H}{i}}} \left( \frac{1+z}{3.4} \right)^{9/2} [\text{cm}^{-2}], 
\label{norm_factor}
\end{equation}
where $\Gamma_{\text{ion},\ion{H}{i}}$ is the photoionization rate of \ion{H}{i} in units of $10^{-12}$~s$^{-1}$
and $T_{4}$ is the temperature of the absorbing gas in units of $10^{4}$~K. 

In the case of simulations, the normalization factor $N_{\ion{H}{i},0} = N_{\ion{H}{i}} (\Delta = 1)$ was calculated using the Eq. (\ref{norm_factor}) for every single thermal model, which was used for calibration (see below). We would like to note that we used the effective UVB $\Gamma_{\text{ion},\ion{H}{i}}$ = $\Gamma_{\text{ion},\ion{H}{i},\text{sim}} / A_r$ as the used simulations were rescaled using the scaling factor $A_{r}$ (see Section \ref{skewer_generation}). By including this step, we reduced the variation of  $N_{\ion{H}{i},0}$ with thermal parameters. In the case of simulation, the average value of $\log{N_{\ion{H}{i},0}} = (12.8247 \pm 0.0872) \rm\, cm^{-2}$ over all of the thermal models was used for the cut-off fitting procedure.

In the case of analysed redshift bins, we calculate the normalization factor according to the equation \citep{Hiss_2018}
\begin{equation}
\log{N_{\ion{H}{i},0}(z)} = 0.6225(1 + z) + 11.1068.
\label{calibrationNHI}
\end{equation}
The reason for this choice is that we used the {\sc THERMAL} suite (see below), which was also used in the aforementioned study to obtain the calibration between $b_{0} - T_{0}$ and $\Gamma - \gamma$.

\subsection{Cut-off fitting results}
For the determination of the final cut-off parameters, we used the bootstrap resampling method with replacement. We generate 2\,000 datasets by bootstrapping of the cut-off fitting procedure, from which the probability distribution functions (PDFs) $p(b_0, \Gamma)$ were obtained. The medians of these distributions was used as the best estimates of the $b_{0}$ and $(\Gamma-1)$ parameters. We would like to note that the uncertainties correspond to the 16th and 84th percentiles of the probability distribution functions. The resulting $b - N_{\ion{H}{i}}$ distributions together with the kernel density estimation of $p(b_0, \Gamma)$ for both studied redshift bins ($1.6 \leq z < 1.8$ and  $1.8 \leq z < 2.0$)  are plotted in Fig. \ref{fig:cutoff_results}.

\section{Simulations}
\label{sec:sim}
To obtained more accurate values of the temperature $T_{0}$ and power-law index $\gamma$, \citet{Bolton2014} suggested that calibration of the cut-off fitting measurements based on the simulations is needed. To reach this goal, we used the set of simulated skewers at redshift $z = 1.8$ from the {\sc THERMAL} (Thermal History and Evolution in Reionization Models of Absorption Lines) \footnote{thermal.joseonorbe.com} suite \citep[for more details see][]{Onorbe2017}. This dataset includes skewers from 88 Nyx hydrodynamical simulations with different combinations of underlying thermal parameters $T_{0}$, $\gamma$ and pressure smoothing scale $\lambda_{P}$ \citep[see details in][]{Onorbe2017, Hiss_2018} on a box size $L_{\rm box} = 20$ Mpc/$h$ and 1024$^3$ cells. The cosmological parameters used in the simulations were based on the results of the \textit{Planck} mission \citep{Planck2014}: $\Omega_{\Lambda} = 0.6808$, $\Omega_{m} = 0.3192$, $\sigma_{8} = 0.826$, $\Omega_{b} = 0.04965$, $n_{s} = 0.9655$, and $h = 0.6704$. The parameters $T_0$ and $\gamma$ were extracted from the simulations by fitting the $T-\rho$ relation power-law to the distribution of gas cells as described in \citet{Lukic2015}. In order to determine $\lambda_{\rm P}$, the cut-off in the power spectrum of the real-space Lyman-$\alpha$ flux was fitted.

\subsection{Skewer generation}
\label{skewer_generation}
Firstly, random Ly-$\alpha$ optical depth $(\tau)$ skewers were transformed into the corresponding flux skewer $F$, i.e. a transmission spectrum along the line of sight, according to the equation $F = F_{c} \exp{(- A_{r} \tau(\lambda))}$, where continuum flux $F_{c}$ was set up to unity and the scaling factor $A_{r}$ allowed the match the lines of sight to observed mean flux values.

Its value was determined by comparing the mean flux of the simulations with observational one, in this case with the value that match the mean flux evolution shown in \citet{Onorbe2017} based on precise measurements by \citet{Fan2006, Becker2007, Kirkman2007, Faucher2008, Becker2013}. We would like to note that we did not consider the impact of uncertainties on the scaling factor value. However, the discussion of mean flux rescaling in the models is presented in Appendix \ref{imp_Flux}.

\subsection{Thermal parameter grid}
\begin{figure}
\centering
	\includegraphics[width=\columnwidth]{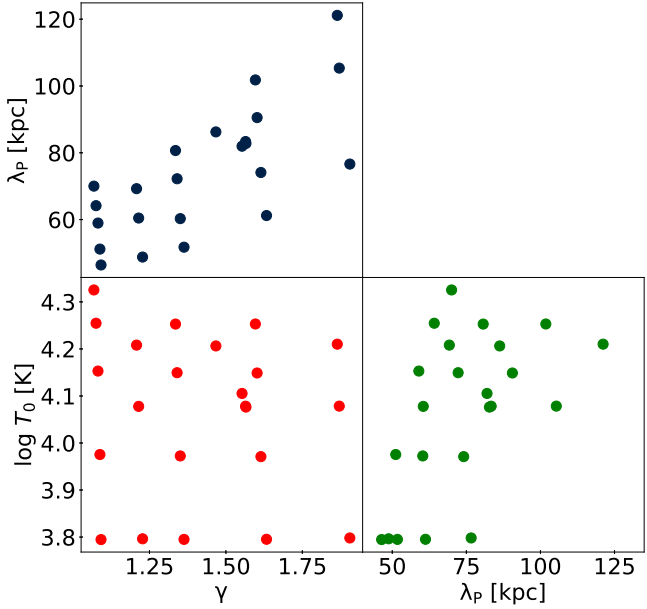}
    \caption{Combinations of parameters $T_{0}$, $\gamma$ and $\lambda_{\rm P}$ used for the calibration procedure.}
    \label{fig:sim_param}
\end{figure}

In this study, we used 23 skewers with a different combination of thermal parameters and pressure smoothing scale. The combinations of parameters $T_{0}, \gamma$ and $\lambda_{\rm P}$ are depicted in the Fig. \ref{fig:sim_param}. The grid of thermal parameters covers intervals: $6\,200 < T_{0} < 21\,200$\,[K] and $1.06 < \gamma < 1.91$. In the case of the pressure smoothing scale, we chose range $ 46 < \lambda_{\rm P} < 121$\,[kpc]. 
\subsection{Modeling Noise, Resolution and Voigt Profile Fitting}

The mock spectra for calibration were prepared by adding the effects of resolution and noise to the simulated skewers. The quantities of both effects were selected to match the used spectra. We mimicked the instrumental resolution by convolving the skewers with a Gaussian with FWHM = 6\,km\,s$^{-1}$ and rebinning to 3\,km\,s$^{-1}$ pixels afterward. In the case of the S/N ratio, we selected the value of 20. We would like to note that the metal contamination was not considered in this case. 

\begin{figure}
\centering
	\includegraphics[width=\columnwidth]{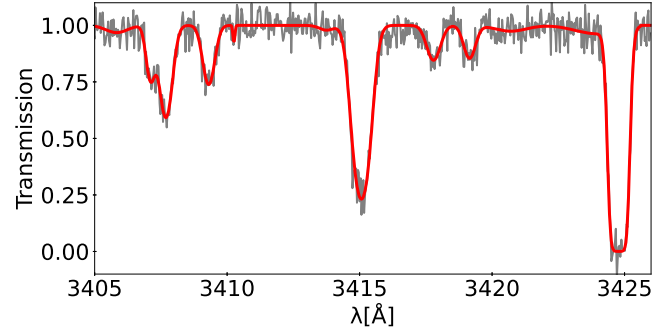}
 \caption{Resulting Voigt profile fit of the Ly-$\alpha$ forest from the simulation with the thermal parameters $\log(T_{0}) = 3.97$ [K], $\gamma = 1.61$,  and smoothing scale $\lambda_{\rm P} = 74$\,kpc. The red line is the spectrum fitted by the {\sc VPFIT} code and the underlying grey solid line is the simulated spectrum.}
    \label{fig:sim_example}
\end{figure}

When the mock spectra were prepared, we used the same Voigt profile fitting procedure as was described in Section \ref{sec:VPFIT}. An example of the decomposition of the Ly-$\alpha$ forest from the simulation into individual absorption lines using {\sc VPFIT} code is shown in Fig. \ref{fig:sim_example}. In this case, we also used the narrow line rejection procedure even if the effect of the rejected lines on the results is neglected. 
\section{Determination of the coefficients in the \boldmath{$T-\rho$} relation}
\label{sec:T_rho_rel}
In this section, we deal with the relationship between the cut-off parameters $b_{0}$, $(\Gamma-1)$ and the thermal parameters $T_{0}$, $\gamma$. Firstly, we rewrote the Eq. (\ref{asymptot_td_relation2}) as
\begin{equation}
\log{T}=\log{T_{0}}+(\gamma-1)\log{\left( \frac{\rho}{\overline{\rho}}\right) },
\label{asymptot_td_relation3}
\end{equation}
where $\rho$ is the baryon density and $\overline{\rho}$ is its mean value ($\rho=\overline{\rho}(1+\delta)$). \citet{Schaye1999} showed that the relations between the density vs temperature and column density vs Doppler parameter of the absorbers near the cut-off can be also fitted by the power laws as
\begin{equation}
\log{\left( \frac{\rho}{\overline{\rho}}\right) }=A+B\log{\left( \frac{N_{\ion{H}{i}}}{N_{\ion{H}{i},0}}\right) },
\label{asymptot_td_relation4}
\end{equation}
\vspace{-0.3cm}
\begin{equation}
\log{T}=C+D\log{b_{\rm th}},
\label{asymptot_td_relation5}
\end{equation}
where the coefficients are given by following equations 
\begin{equation}
\log{b_{0}}=\frac{1}{D}[\log{T_{0}}-C+(\gamma-1)A],
\label{asymptot_td_relation7}
\end{equation}
\vspace{-0.3cm}
\begin{equation}
(\Gamma-1)=\frac{B}{D}(\gamma-1).
\label{asymptot_td_relation8}
\end{equation}
It is worth noting that the Eq. (\ref{cutoff1}) characterizes the minimal broadening at a given $N_{\ion{H}{i}}$, where the absorbers are thermally broadened (therefore $b_{\rm th}$ in the relation). The chosen value of $N_{\ion{H}{i},0}$ represents the column density of a cloud with mean density, therefore $A = 0$, and the dependency on $\gamma$ disappears from the Eq. (\ref{asymptot_td_relation7}) \citep{Hiss_2018}. Based on these assumptions, we can rewrite the Eqs. (\ref{asymptot_td_relation7}) and (\ref{asymptot_td_relation8}) as
\begin{equation}
\log{T_{0}}= C + D \log{b_{0}},
\label{asymptot_td_relation9}
\end{equation}
\vspace{-0.5cm}
\begin{equation}
(\gamma-1)=\kappa(\Gamma-1),
\label{asymptot_td_relation10}
\end{equation}
where $\kappa=D/B$.

\subsection{Calibration Using Simulations}
\begin{figure}
\centering
	\includegraphics[width=\columnwidth]{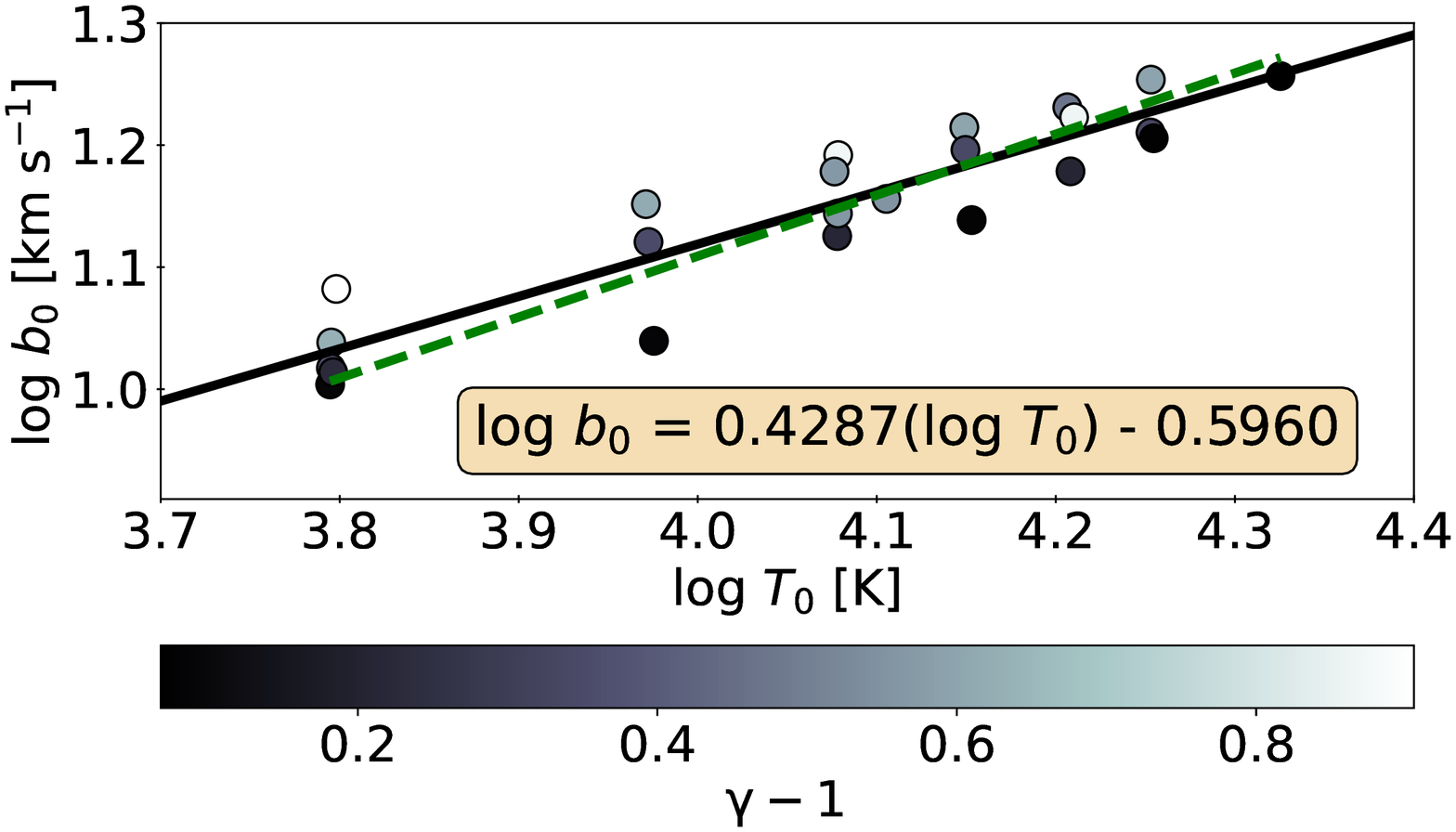}\\
	\vspace{2em}
	\includegraphics[width=\columnwidth]{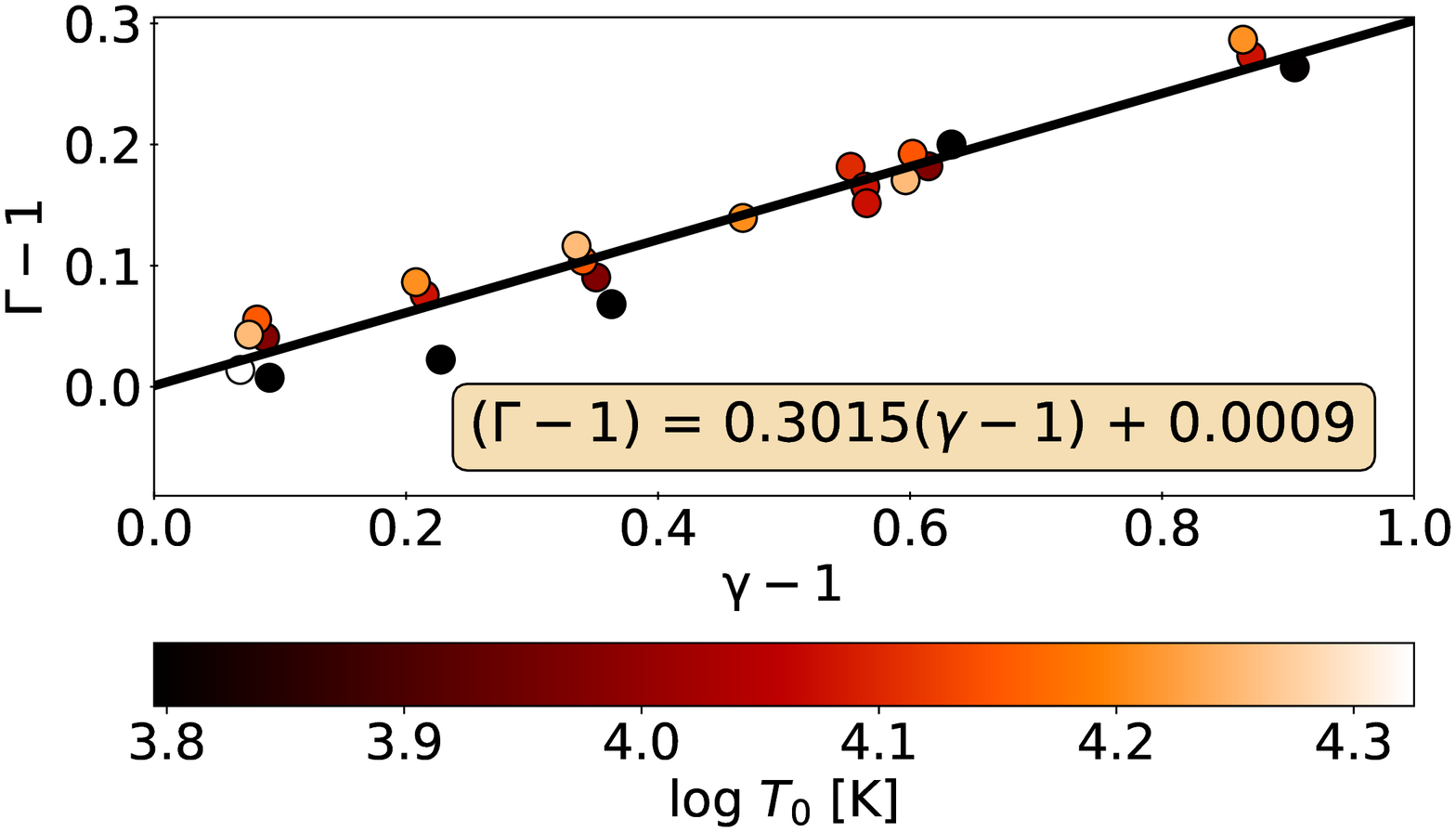}
    \caption{Calibrations of the $\log{b_0}$ vs $\log{T_{0}}$ (upper panel) and $(\Gamma - 1)$ vs $(\gamma - 1)$ (bottom panel) relations. The black lines correspond to the best linear fits to the points. The case, when the value of $b_{0}$ corresponds to pure thermal broadening is depicted by the green dash line.}
    \label{fig:Calibration_Graph}
\end{figure}

To generate the calibrations between $b_{0} - T_{0}$ and $\Gamma - \gamma$, we used 23 selected thermal models at the redshift $z = 1.8$. In this case, the cut-off fitting algorithm was used on simulated $b - N_{\ion{H}{i}}$ distributions, each of which was constructed using 120 mock spectra from all used thermal models. The results are shown in Fig. \ref{fig:Calibration_Graph}, where depicted points correspond to the median values of $b_{0}$ and $\Gamma - 1$ from 2\,000 datasets generated by bootstrapping. This is the same approach as we used for the observational data.

To include the additional effects, such as various values of $\lambda_{\rm P}$ and using of the same value of the $\log{N_{\ion{H}{i}}}$ for all models, we fitted Eqs. (\ref{asymptot_td_relation9}) and (\ref{asymptot_td_relation10}) to the 2\,000 bootstrap realizations of the points in the $\log{(T_{0})} - \log{(b_{o})}$ and $(\gamma - 1) - (\Gamma - 1)$ diagrams (see Fig. \ref{fig:Calibration_Graph}) with replacements. In the $\log{(b_{0})}$ vs $\log{(T_{0})}$ diagram, we also included the case when the value of $b_{0}$ is caused purely by thermal broadening, i.e. $b_{0} = \sqrt{2 k_{\rm B} T_{0} / m_{p}}$, where $k_{\rm B}$ is the Boltzmann constant and $m_{p}$ is the proton mass.

The best fit values of the coefficients $C$, $D$ and $\kappa$ correspond to the medians of bootstrap distributions. Using this approach, we obtained the values: $C = 1.39^{+0.20}_{-0.24}$, $D = 2.33^{+0.20}_{-0.17}$, and $\kappa = 3.32^{+0.17}_{-0.17}$, where the uncertainties correspond to the 16th and 84th percentiles of the $p(D,C)$ and $p(\kappa)$.
\section{Results and discussion}
\label{sec:results}
The $b - N_{\ion{H}{i}}$ distributions with the corresponding cut-off fits and PDFs $p(b_{0}, \Gamma)$ for both studied redshift bins ($1.6 \leq z < 1.8$ and  \mbox{$1.8 \leq z < 2.0$)} are depicted in Fig. \ref{fig:cutoff_results}. The PDFs indicate the anticorrelation between $b_{0}$ and $(\Gamma-1)$. 

\begin{figure*}
\centering
	\includegraphics[width=0.8\columnwidth]{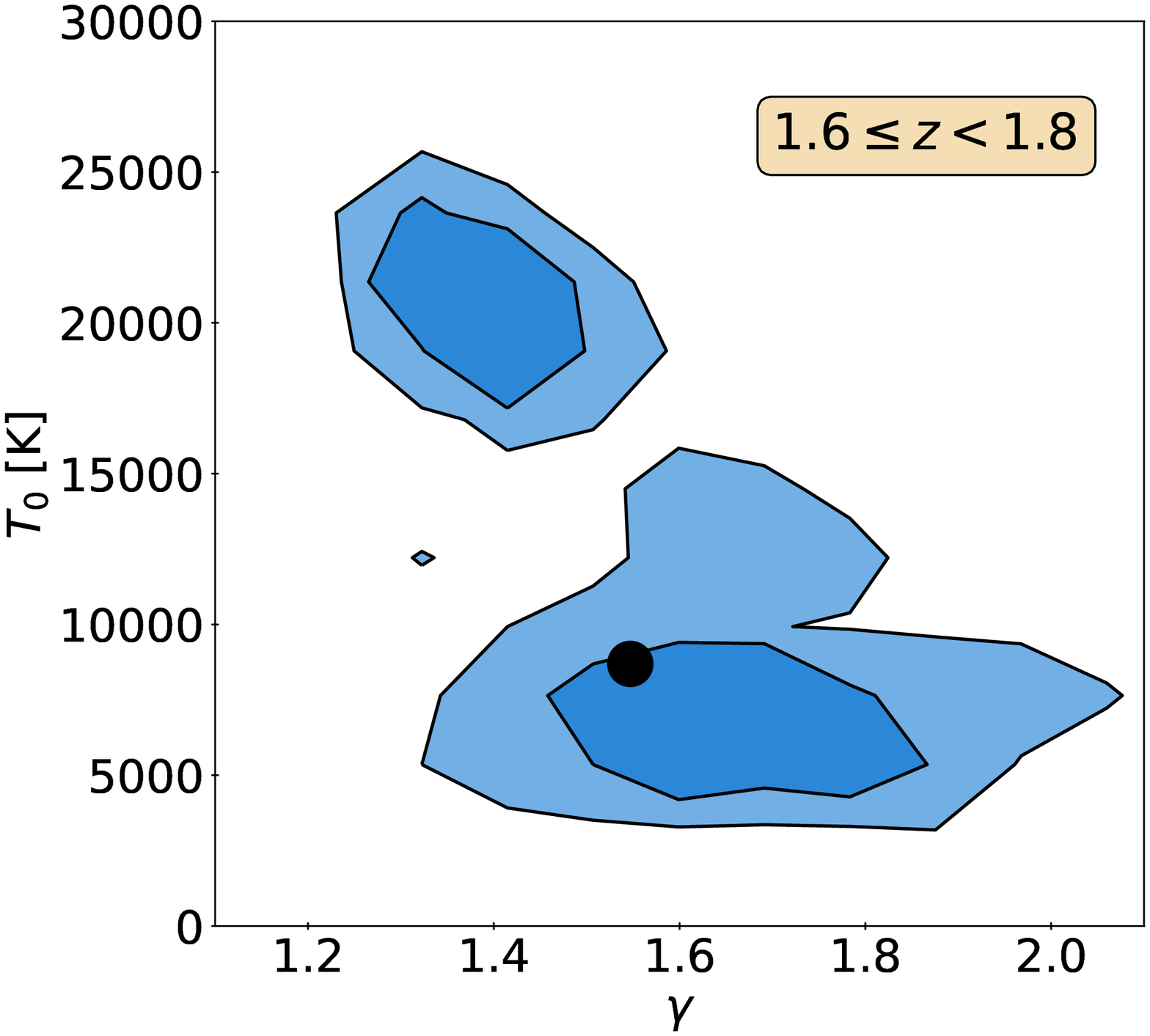} \hspace{3em}
	\includegraphics[width=0.8\columnwidth]{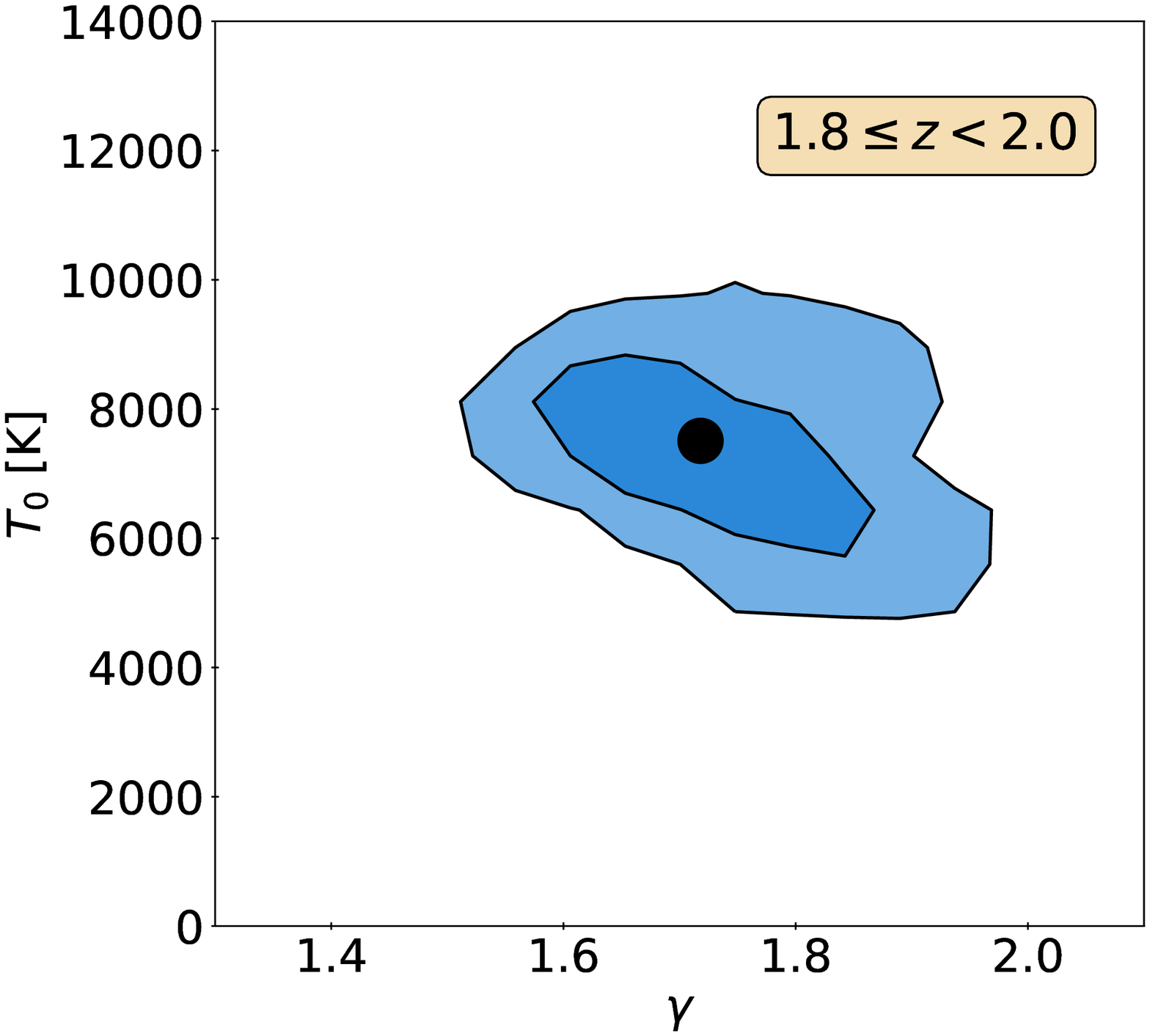}
	\caption{Resulting PDFs $p(T_{0}, \gamma)$ for the redshift range $1.6 \leq z < 1.8$ (left panel) and  $1.8 \leq z < 2.0$ (right panel). The 68\% and 95\% confidence levels are plotted by dark and light blue color, respectively. The black points correspond to the median of the marginal distributions of $T_{0}$ and $\gamma$.}
    \label{fig:PDFs_final}
\end{figure*}

To include uncertainties which arose during the individual steps implemented in the analysis, we used the approach described by \citet{Hiss_2018}. We combined the 2\,000 bootstrapped $b_{0}$ and $(\Gamma-1)$ pairs in $p(b_{0}, \Gamma)$ with every point in the boostrapped calibration PDFs $p(D, C)$ and $p(\kappa)$ from simulations using the Eqs. (\ref{asymptot_td_relation9}) and (\ref{asymptot_td_relation10}). The resulting PDFs $p(T_{0}, \gamma)$ for both studied redshift intervals are depicted in Fig. \ref{fig:PDFs_final}. The medians of bootstrap distributions were used as the best estimates of the $T_{0}$ and $\gamma$. The derived parameter values for the redshift intervals $1.6 \leq z < 1.8$ and  $1.8 \leq z < 2.0$  are listed in Tab \ref{tab:Results}. The uncertainties correspond to the 16th and 84th percentiles of the probability distribution function $p(T_{0}, \gamma)$. The comparison of the results obtained in this study with previously published ones is shown in Fig. \ref{fig:models_comp1}.

\cite{Schaye1999} showed that the effect of the S/N ratio on the results is negligible if its value is higher than $\sim 20$. However, the median of this parameter for the spectra used in this study is equal to 18. We tested the possibility that the lower value of $T_{0}$ and higher value of $\gamma$ for mean value of redshift $z = 1.8$ could be affected by the lower S/N ratio value, but our analysis showed that this effect is possible to neglect (see Appendix \ref{SN_effect}).

We also considered the possibility that the obtained values of the calibration parameters could be influenced by using the calibration for the redshift $z = 1.8$, whereas the centers of the redshift bins analysed in this study are 1.7 and 1.9. Based on the dependencies of the calibration parameters $C$, $D$, and $\kappa$ on the redshift presented by \citet{Hiss_2018} (Fig. 14 therein), we can conclude that changes of their values for the redshift differences $\Delta z = 0.1$ are significantly lower than their uncertainties determined in this study, and this effect can be considered negligible. 

%%Nase vysledky pre z=1.8 su nasledovne: $D = 2.33^{+0.20}_{-0.17}$, $C = 1.39^{+0.20}_{-0.24}$ and $\kappa = 3.32^{+0.17}_{-0.17}$ (section 4.1). V praci od Hiss et al. v grafe 14 je pre ilustraciu uvedena zavislost parametrov C, D a \kappa pre 2 < z 3.4, kde rozdiel pre \Delta z je v pripade D = 0.06, C= - 0.08 a kappa = -0.06 (Tieto hodnoty su iba orientacne, ale su nizsie nez neistoty nasich hodnot)

The extended uncertainties of the derived parameters for the redshift range $1.6 \leq z < 1.8$ originate from the multimodal distribution $p(b_{0}, \Gamma)$ (see Figs. \ref{fig:cutoff_results}). It is worth noting that the simulated $b - N_{\ion{H}{i}}$ distributions led to the unimodal solutions. Therefore, this behavior can be attributed to some systematics related with the used spectra, for which the cut-off fitting led to the artifacts. We assume that the effect is caused by the number of absorbers ($\sim 550$) used for this redshift bin.
\begin{table}
	\centering
	\caption{Summary of the derived parameters for the studied redshift bins.}
	\label{tab:Results}
	\begin{tabular}{ccccc} 
		\hline
		\textbf{Redshift} & \boldmath{{$b_{0}$}} & \boldmath{{$T_{0}$}} & \boldmath{{$\Gamma$}} & \boldmath{{$\gamma$}} \\
		\textbf{range} &\boldmath{$\rm [km\,s^{-1}$]} & \boldmath{$[10^{3}\rm\,K$]} & & \\
		\hline
$\langle 1.6, 1.8)$ & $12.26^{+5.67}_{-1.49}$ & $8.69^{+12.05}_{-2.48}$ & $1.17^{+0.05}_{-0.06}$ & $1.55^{+0.20}_{-0.18}$\\[0.2cm]
$\langle 1.8, 2.0)$ & $11.78^{+0.29}_{-0.83}$ & $7.51^{+0.81}_{-1.09}$ & $1.22^{+0.03}_{-0.02}$ & $1.72^{+0.11}_{-0.08}$\\
		\hline
	\end{tabular}
\end{table}

\subsection{Comparison with previous studies}
By comparing the values of $T_{0}$ and $\gamma$ obtained in this study with previously published ones (Fig. \ref{fig:models_comp1}), it can be concluded that even there are some discrepancies, the results correspond to each other. An exception is the study by \cite{Boera2014}, in which the temperature at the mean density assuming the value of $\gamma \sim 1.3$ is significantly higher for both mean redshifts ($z = 1.63$ and $1.82$) even compared to simulations. The differences can be attributed to the different methods used to estimate the values of $T_{0}$ and $\gamma$.

Only \citet{Schaye2000} studied the IGM in the redshift interval $1.85 \leq z \leq 2.09$ using the Voigt profile decomposition of the Ly-$\alpha$ forest into the set of individual absorption lines, as was used in this work. For the median value of the redshift $z = 1.96$, the authors determined value of $T_{0} \sim 11\,000$\,K and $\gamma = 1.4$, which differ from the ones obtained in our study for the redshift range $\langle1.8, 2.0)$. It is worth noting that \citet{Schaye2000} used the spectrum of only one quasar (Q1100-264), the different value of $\log{N_{\ion{H}{i},0}} = 13$, and their median redshift value was slightly different from the one in this study ($z = 1.9$).

\subsection{Comparison with models of the IGM thermal evolution}
\begin{figure*}
\centering
	\includegraphics[width=0.8\textwidth]{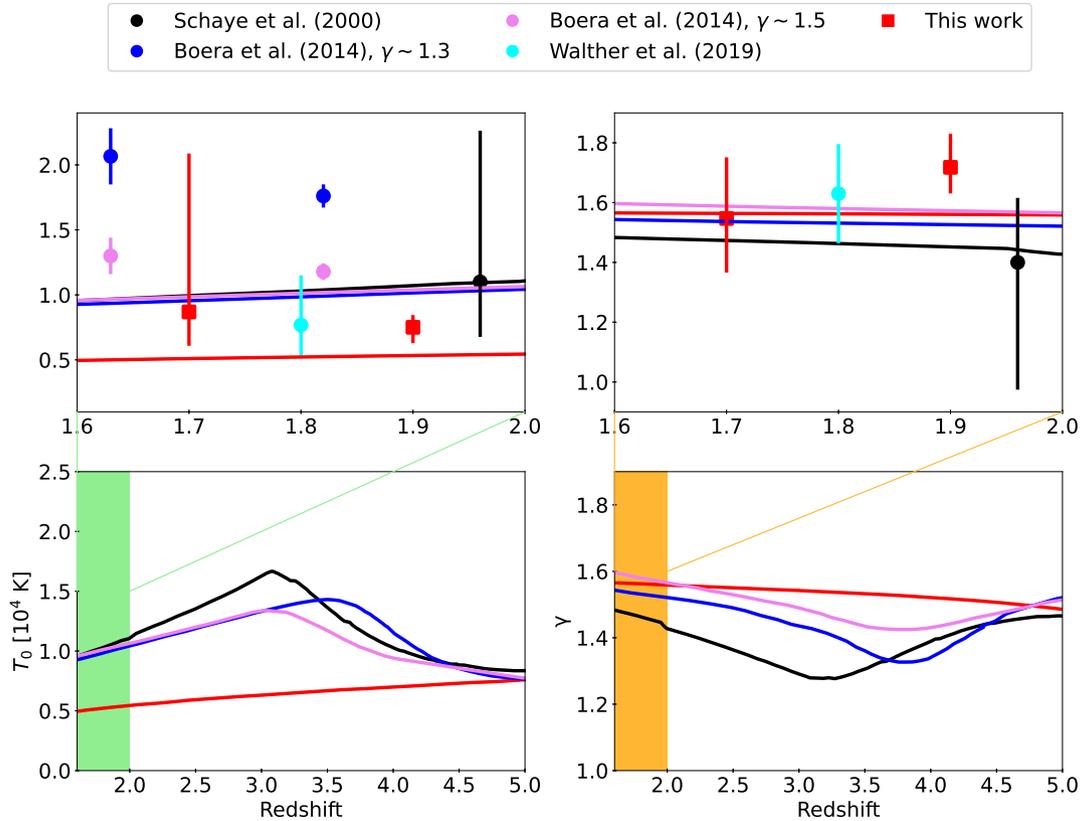}
    \caption{Comparison of the results obtained in this study with previously published ones and various models. The black and red curves correspond to the semi-analytical models by \citet{Upton2016} where the He\,{\sc II} reionization has and has not occurred, respectively. The blue and violet curves correspond to the non-equilibrium simulation by \citet{Puchwein_2015} and the hydrodynamical simulation by \citet{Onorbe2017}, respectively.}
    \label{fig:models_comp1}
\end{figure*}

We compared the values of $T_{0}$ and $\gamma$ obtained in this study with results of various models of the IGM thermal evolution (see Fig. \ref{fig:models_comp1}). Namely, the non-equilibrium simulation by \citet{Puchwein_2015}, the semi-analytical models by \citet{Upton2016} where the He\,{\sc II} reionization has and has not occurred, and hydrodynamical simulation by \citet{Onorbe2017}. The semi-analytical models by \citet{Upton2016} were used only for qualitative comparison as these models depend on a number of parameters such as the ionizing background spectral index, the quasar spectral index, duration of the reionization, clumping etc. In addition, these models are not independent of observational data because the authors used the temperature measurements of \citet{Becker2011} and \citet{Boera2014} for constraining the model curves.

The comparison shows that the obtained temperatures at mean density for both redshift bins $\langle1.6, 1.8)$ and $\langle1.8, 2.0)$ are slightly lower than predicted by various models. 
In the case of redshift bin $\langle1.6, 1.8)$, the value of $T_0$ corresponds to the simulations within its error, although the uncertainty is considerably large. 

The same conclusion can also be stated in the case of the power-law index, for which its value for the  redshift bin $\langle1.6, 1.8)$ matches the non-equilibrium simulation of \citet{Puchwein_2015}, and within the error corresponds to all presented models. On the other hand, for the redshift bin $\langle1.8, 2.0)$, we obtained the value of the power-law index which is higher than the expected one ($\gamma \sim1.6$) resulting from the balance of photoheating and adiabatic cooling after all reionization events \citep{HuiGnedin, Upton2016}.

\section{Conclusions}
\label{sec:conclusions}
In this study, the sample of 35 quasar spectra was used for inferring parameters of the temperature-density relation in the IGM for the redshift interval $\langle1.6, 2.0)$. For this purpose, we used the approach that treats the Ly-$\alpha$ forest as a superposition of discrete absorption profiles. Based on the fitting, the $b-N_{\ion{H}{i}}$ distributions were constructed, from which the position of thermal state sensitive cut-off was determined. We also calibrate our measurements using 23 combinations of thermal parameters and pressure smoothing scale from the {\sc THERMAL} suite of hydrodynamical simulations. The main results can be summarized as follows:
\begin{itemize}
\item The obtained values correspond to the decreasing trend of the temperature at mean density predicted by the various models toward the lower redshifts.
\item For the redshift interval $\langle1.6, 1.8)$, we obtained the power law index $\gamma = 1.55$. The value is close to $1.6$, as expected by various models, and results from the balance of photoheating with adiabatic cooling after all reionization events. However, the determined value $\gamma = 1.72$ is higher for the redshift interval $\langle1.8, 2.0)$.
\item Our determinations of $T_{0}$ and $\gamma$ at the lower redshift interval $\langle1.6, 2.0)$ indicate the completion of the reheating process associated to \ion{He}{ii} reionization. 
\end{itemize}

\section*{Acknowledgements}
This research is based on the use of the Keck Observatory Archive (KOA), which is operated by the W. M. Keck Observatory and the NASA Exoplanet Science Institute (NExScI), under contract with the National Aeronautics and Space Administration, and also on data products created from observations collected at the European Organisation for Astronomical Research in the Southern Hemisphere. The authors would also like to thank Jose O\~{n}orbe for fruitful discussion.

\section*{Data Availability}
The data underlying this article will be shared on reasonable request to the corresponding author.

\bibliographystyle{mnras}
\bibliography{bibliography}
% if your bibtex file is called example.bib

%%%%%%%%%%%%%%%%% APPENDICES %%%%%%%%%%%%%%%%%%%%%

\appendix
\section{Effects of continuum misplacement}
\label{continuum_misplacement}
In this section, we discuss the effect of continuum misplacement in used data. In the case of high-resolution, high S/N spectra, the continuum is fitted locally by connecting apparent absorption-free spectral intervals \citep{McDonald2000,Kim2007}. As was noted in \citet{Kim2007}, this method is applicable for the redshift $1.5 < z_{\text{em}} < 3.5$, where the continuum placement statistical uncertainty is of the order of a few percent.

As the continuum placement affects the corresponding optical depth of the spectral line, and consequently the line profile parameters, we address the effect by calculating the optical depth $\tau$ in the line center as \citep{Padmanabhan_Theor_Astro_3}

\begin{equation}
\tau_{\text{lc}}({\ion{H}{i}}) = 1.497\times10^{-15} \frac{N f \lambda_0}{b},
\label{tau_center}
\end{equation}

where $\lambda_{0} = 1215.67$\,\AA,\,\, $N\,[\rm cm^{-2}]$ and $b\,[\rm km\,s^{-1}]$. For the calculation $\tau_{\text{lc}}({\ion{H}{i}})$ using Eq. (\ref{tau_center}), we used the typical value of the Doppler parameter in our sample $b = 13\,\rm km\,s^{-1}$ for spectral lines with various column densities. Then, we calculated the flux at line center according to equation $F_{\text{lc}} = F_{\text{c}} \exp{(-\tau_{\text{lc}}({\ion{H}{i}}))}$, where $F_{\text{c}}$ was set up to unity, and the value of $F_{\text{lc}}$ was shifted to mimic the effect of misplacement of the continuum. We assumed that our typical continuum uncertainty is $\sim$5\%. The final step was to calculate the corresponding column density by the reverse calculation with the fixed value of Doppler parameter. 

The results showed that for the continuum shift of 5\%, the corresponding shift of $\log{N_{\ion{H}{i}}}$ is comparable to the {\sc VPFIT} uncertainty at $\log{N_{\ion{H}{i}}} \approx 12.5$ and exceeds it at lower values of column densities. From this result we can concluded that the effect of the continuum misplacement is possible to neglect within the cut-off fitting range used in this study.

\section{Mean flux rescaling in the models}
\label{imp_Flux}

\begin{figure}
\centering
	\includegraphics[width=\columnwidth]{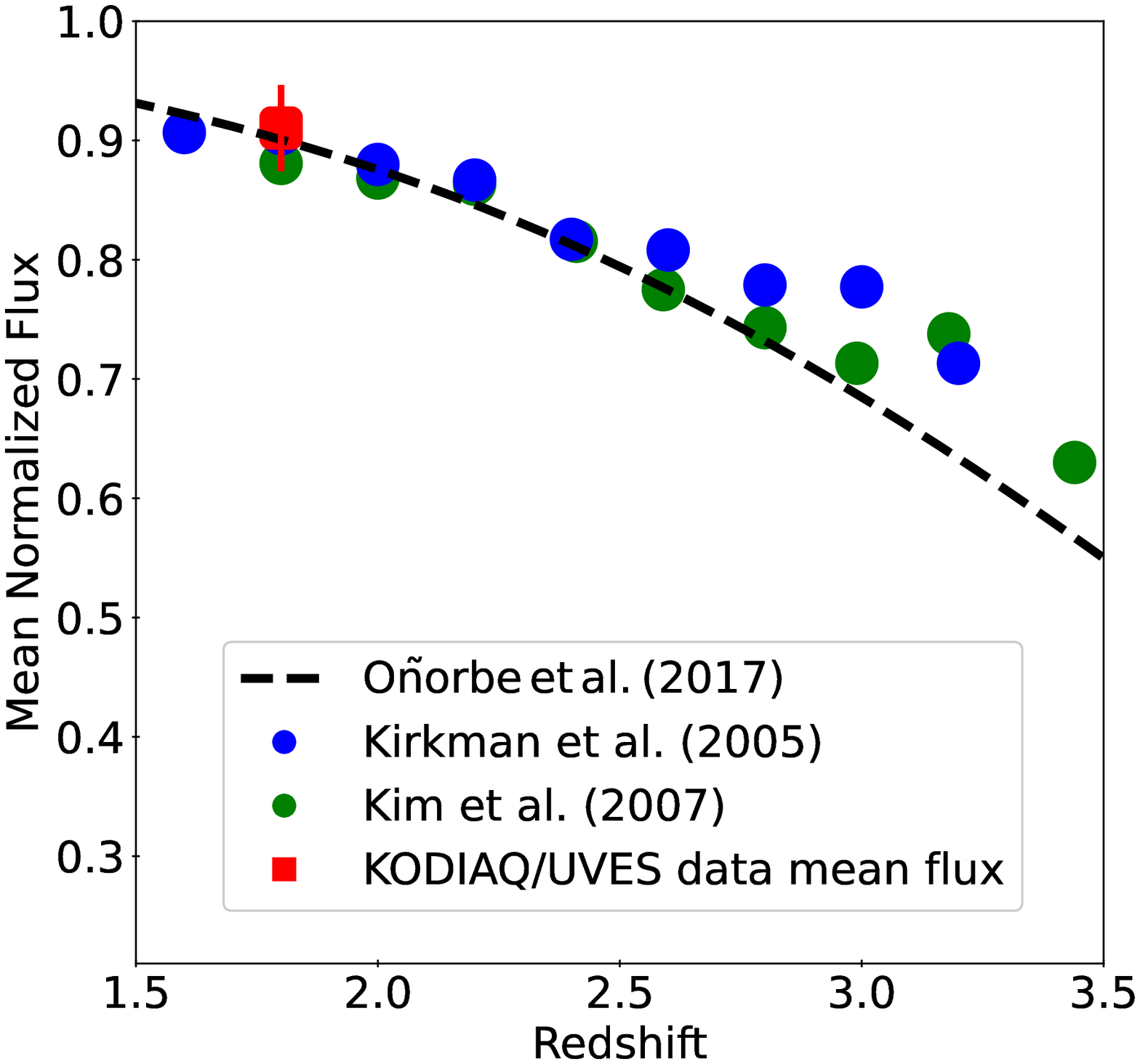}
    \caption{The comparison of the mean flux evolutions from various studies with the mean flux of the data used in this work.}
    \label{fig:mean_flux}
\end{figure}

As noted in Section \ref{skewer_generation}, we rescaled the fluxes obtained from the simulations in order to match the observed mean flux $\overline{F}$. In this work, we chose the value corresponding the mean flux evolution in \citet{Onorbe2017}, which is based on precise measurements by \citet{Fan2006, Becker2007, Kirkman2007, Faucher2008, Becker2013}. The $\overline{F}$ value used in the case of simulations corresponds to the mean transmitted flux determined based on the KODIAQ/UVES spectra. The comparison is plotted in Fig. \ref{fig:mean_flux} together with the values determined from the studies of \cite{Kirkman2005} and \cite{Kim2007}. We direct the reader to the study of \citet{Hiss_2018} for the discussion of the effects of mean flux rescaling in the models. 

\section{Effect of the S/N ratio on the results of cut-off procedure}
\label{SN_effect}
\begin{figure}
\centering
	\includegraphics[width=0.8\columnwidth]{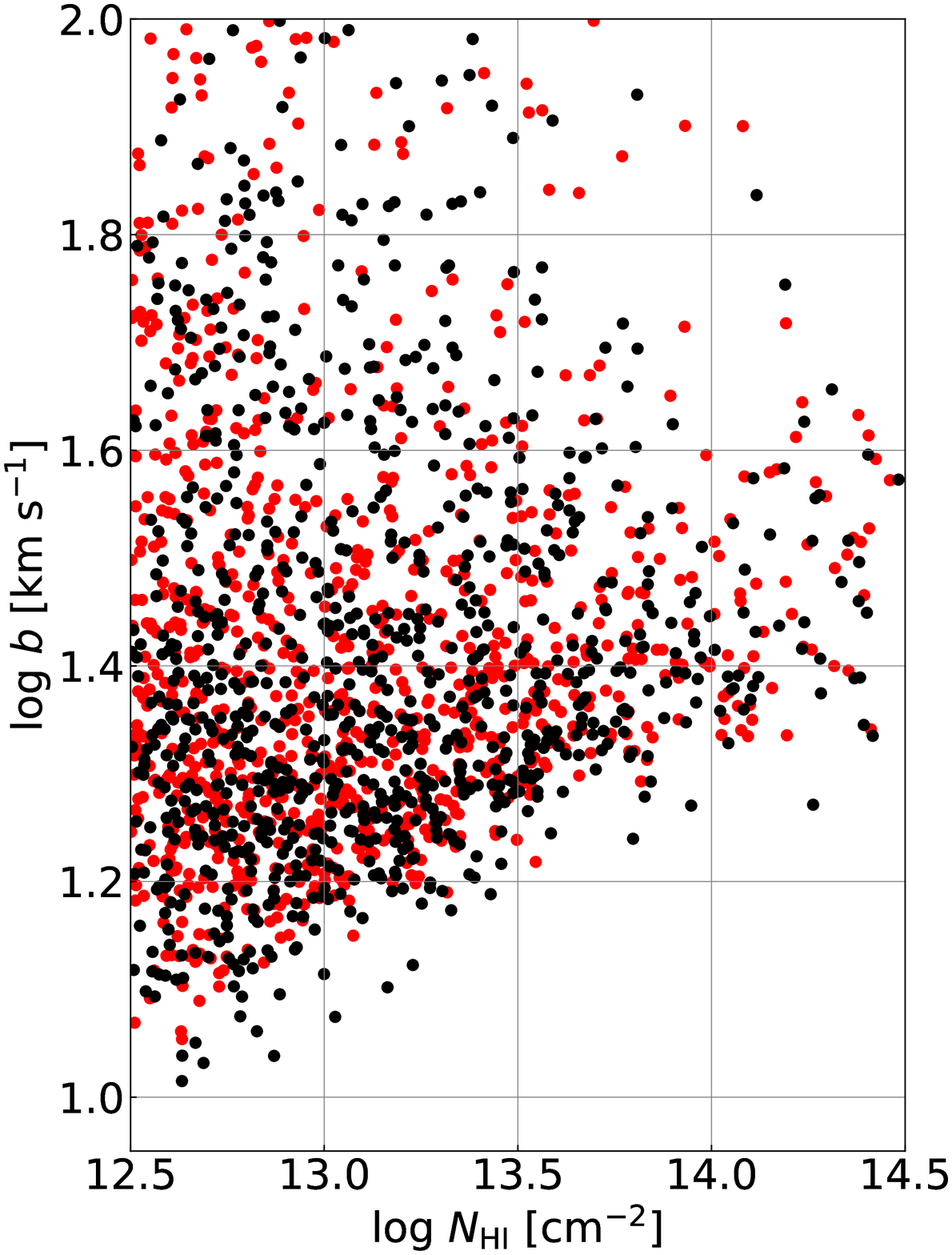}
    \caption{Simulated $b-N_{\ion{H}{i}}$ distributions at $z = 1.8$ with various S/N ratio values applied to lines of sight. The distributions were generated by Voigt profile fitting the same skewers as were used in the case of calibration process. The red points depict the distribution with S/N = 83 which corresponds to the highest S/N ratio value of used spectra. The black points correspond to the distribution with S/N = 20 characterizing the S/N ratio used in the case of the simulations. The thermal parameters used in these mock distributions are $\log(T_{0}) = 3.97$ [K], $\gamma = 1.61$, and smoothing scale $\lambda_{\rm P} = 74$\,kpc.}
    \label{fig:SN_comp}
\end{figure}

To test the effects of the S/N ratio for values $\lesssim 20$ \citep{Schaye1999}, we simulated two $b-N_{\ion{H}{i}}$ distributions at $z = 1.8$ with different S/N ratios applied to lines of sight (see Fig. \ref{fig:SN_comp}). The first one corresponds to the highest S/N value of used spectra (83), and the second one corresponds to the value, which was used in the case of simulations (20). After application of the same cut-off fitting procedure as was used in the case of spectra and simulations, we obtained the values $b_{0} = 14.18^{+0.30}_{-0.38}$, $(\Gamma - 1) = 0.18^{+0.01}_{-0.02}$ and \mbox{$b_{0} = 14.31^{+0.32}_{-0.67}$}, \mbox{$(\Gamma - 1) = 0.17^{+0.01}_{-0.01}$} for the S/N = 20 and \mbox{S/N = 83}, respectively. Since the values of the parameters are same within their errors, we can conclude that the effect of the S/N ratio is negligible.

% Don't change these lines
\bsp	% typesetting comment
\label{lastpage}
\end{document}